# Wavelength dependent optical enhancement of superconducting interlayer coupling in La$_{1.885}$Ba$_{0.115}$CuO$_4$

E. Casandruc[1,2], D. Nicoletti[1,2], S. Rajasekaran[1,2], Y. Laplace[1,2], V. Khanna[1,2,4,5], G. D. Gu[3], J. P. Hill[3], and A. Cavalleri[1,2,4]

[1] *Max Planck Institute for the Structure and Dynamics of Matter, Hamburg, Germany*

[2] *Center for Free Electron Laser Science, Hamburg, Germany*

[3] *Condensed Matter Physics and Materials Science Department, Brookhaven National Laboratory, Upton, NY, United States*

[4] *Department of Physics, Clarendon Laboratory, University of Oxford, Oxford, United Kingdom*

[5] *Diamond Light Source, Chilton, Didcot, Oxfordshire, United Kingdom*

**Abstract**

We analyze the pump wavelength dependence for the photo-induced enhancement of interlayer coupling in La$_{1.885}$Ba$_{0.115}$CuO$_4$, which is promoted by optical melting of the stripe order. In the equilibrium superconducting state ($T < T_C$ = 13 K), in which stripes and superconductivity coexist, time-domain THz spectroscopy reveals a photo-induced blue-shift of the Josephson Plasma Resonance after excitation with optical pulses polarized perpendicular to the CuO$_2$ planes. In the striped, non-superconducting state ($T_C < T < T_{SO} \simeq 40$ K) a transient plasma resonance similar to that seen below $T_C$ appears from a featureless equilibrium reflectivity. Most strikingly, both these effects become stronger upon tuning of the pump wavelength from the mid-infrared to the visible, underscoring an unconventional competition between stripe order and superconductivity, which occurs on energy scales far above the ordering temperature.

Many materials in the cuprate family show spin- and charge-density-wave orders[1,2,3,4,5,6,7], whose interaction with superconductivity is not well understood. $La_{2-x}Ba_xCuO_4$ (LBCO) is a prototypical high-$T_C$ "striped" superconductor, displaying charge and spin modulations that suppress superconductivity in a narrow doping range around $x$ = 1/8 (see phase diagram in Fig. 1a). Along with this so-called "1/8 anomaly" one finds also a structural transition from a low-temperature orthorhombic (LTO) to a low-temperature tetragonal (LTT) lattice symmetry, which is known to stabilize the stripe order[8,9].

Recently, it was shown that superconducting interlayer coupling can be transiently enhanced for doping values below the 1/8 anomaly by excitation with near-infrared pulses polarized perpendicular to the $CuO_2$ planes[10]. This effect is different from the enhancement obtained by phonon excitation[11,12,13], and descends from the weakening of stripe order by charge excitation[14] rather than by a deformation of the lattice[15,16,17]. In contrast to the response with light polarized along the $CuO_2$ planes[18,19,20,21,22,23], $c$-axis optical excitation preferentially weakens stripe order and not the superconductor[24]. Only at later time delays decoherence of the enhanced superconducting state sets in, presumably due to quasi-particle relaxation[10].

Here, we study the pump photon energy dependence of the enhanced superconducting coupling by tuning the pump wavelength between the mid-infrared (5 μm) and the visible (400 nm). We focus our analysis on the $x$ = 11.5% compound, which shows the most striking response of all doping levels studied[10]. In this material, superconductivity, spin and charge order appear at $T_C$ = 13 K, $T_{SO} \simeq 41$ K and $T_{CO} \simeq 53$ K respectively. The sample was a single crystal grown with the floating zone technique[25], which was then cut and polished to give an $ac$ surface[10] of

~5 mm². The excitation was performed with femtosecond optical pulses polarized perpendicular to the $CuO_2$ planes, tuned to central wavelengths of 5 μm, 2 μm, 800 nm, and 400 nm. The experiments were carried out at two temperatures below and above $T_C$, T = 4 K ($T < T_C$) and T = 30 K ($T_C < T < T_{SO}$), marked as dots in the phase diagram in Fig. 1a. Nonlinear conversion of 800-nm wavelength pulses from a Ti:Sa laser was used to generate tunable pump pulses, with a combination of frequency doubling ($\lambda_{pump}$ = 400 nm), optical parametric amplification ($\lambda_{pump}$ = 2 μm) and difference frequency generation between signal and idler pulses ($\lambda_{pump}$ = 5 μm). The pump photon energies are indicated with arrows in Fig. 1b.

The optically-induced dynamics was probed with single-cycle THz probe pulses, generated by illuminating a photoconductive antenna with a replica of the near-infrared laser pulses used for excitation. The probe THz field was also polarized perpendicular to the $CuO_2$ planes and detected the time dependent interlayer coupling strength after photo-excitation. The frequency-resolved transient response was measured between ~150 GHz and ~3 THz by electro-optical sampling of the THz field reflected from the sample with a second replica of the 800-nm Ti:Sa pulses in a ZnTe crystal.

The transient optical response was obtained as a function of pump-probe delay by measuring the pump-induced change in reflected THz electric field. The equilibrium optical properties, to which these transient spectra were referenced, were obtained[26] by comparing the THz reflected field measured at equilibrium with our setup at different temperatures against the *c*-axis broadband reflectivity reported in Ref. 27 (see broadband equilibrium spectra in Fig. 1b). The transient response was then processed by taking into account the mismatch between the penetration depth of the pump pulses (~0.1 - 10 µm, depending on the excitation wavelength) and the THz probe pulses (~50 - 500 µm). This was achieved by assuming a photo-excited layer and an unperturbed bulk volume beneath it[28], as also described in detail in Refs. 29 and 30.

In figure 2, we report the reflectivity changes in the photo-excited layer induced by pump pulses of different wavelengths, 1.5 ps after excitation. For excitation at 5 µm, no appreciable pump-induced effects were detected at any temperature (panels a.1 and a.2). As the pump pulses were tuned to shorter wavelengths, a signal of increasing strength emerged. For 2-µm wavelength pump pulses, a shift of the equilibrium Josephson plasma resonance (JPR) toward higher frequencies was observed below $T_C$ (panel b.1). At 30 K (T > $T_C$) a reflectivity edge appeared at ~200 GHz from the featureless equilibrium reflectivity (black curve in panel b.2). Qualitatively similar but stronger effects were observed for 800-nm optical excitation, with a striking shift of the equilibrium JPR from ~200 GHz to ~600 GHz (panel c.1) for $T < T_C$ and a transient edge near 500 GHz for $T_C < T < T_{SO}$ (panel c.2). Finally, in the case of 400-nm excitation, although a strong increase in the sample reflectivity was observed (panels d.1-d.2), no sharp edge was found.

Two opposing trends can be identified when tuning the pump wavelength, both below and above $T_C$. The transient photo-induced reflectivity edge becomes more pronounced and appears at progressively higher energies for shorter pump wavelengths. However, the "quality" of such edge, as identified by its size and width, also deteriorates for higher photon energy excitation, indicating that decoherence becomes progressively more important.

To analyze the origin of these observations quantitatively, we first note that linear optical absorption in this compound increases with photon energy (as seen in the optical conductivity in figure 1b). Hence, for a given fluence, the total energy and the number of photons deposited in a unit volume varies with the wavelength of the pump pulses. Figure 3 shows the fluence dependent spectrally integrated response (measured as the change in the THz electric field peak $\Delta E_R/E_R$ at a pump-probe delay $\tau$ = 1.5 ps) for three different wavelengths. The signal was found to saturate in all cases with fluence, although this occurred anywhere between ~1 mJ/cm$^2$ and ~3 mJ/cm$^2$. However, when renormalized against the total number of absorbed photons per unit volume (see upper scale for each panel), we found that the optical response always saturated for ~10$^{20}$ photons/cm$^3$. Hence, the different blue shifts and edge widths, which are measured in the saturated regime of figure 3 for comparable absorbed photon numbers, are to be interpreted as an intrinsic feature of stripe melting, which is not dependent on the different excitation conditions.

The two competing phenomena of enhanced coupling (blue shift of the plasma edge) and increased width of the resonance (decoherence) can be better seen when analyzing the transient complex optical conductivity $\sigma_1(\omega,\tau) + i\sigma_2(\omega,\tau)$ of the photo-excited material, where $\omega$ is the probe frequency and $\tau$ is the time delay with respect to the pump pulse. We first discuss the complex response at one time delay immediately after excitation ($\tau$ = 1.5 ps). Representative conductivities at this time delay are shown in figure 4. As already discussed for the reflectivity response of figure 2, no effect was observed for 5-µm pump pulses (panels a.1-4). In contrast, for excitation with 2-µm pulses, we measured an increase in the imaginary (inductive) part of the conductivity, $\sigma_2(\omega)$, which was observed to become positive and increase with decreasing frequency.

An increase of a *positive* $\sigma_2(\omega)$ with decreasing frequency is connected with perfect transport and is typically taken as indicative of a transient superconducting response. Although $\sigma_2(\omega)$ scales as $1/\omega$ in an ideal superconductor, in the cuprates the imaginary conductivity is almost linear due to combined effect of the condensate and of quasiparticle tunneling[30]. The increase in the inductive response is initially accompanied by a negligible change of the ohmic conductivity $\sigma_1(\omega)$, indicating small quasi-particle heating. A qualitatively similar response is reported for the striped state above $T_C$.

For 800-nm wavelength excitation the same effect emerges even more clearly, with an enhancement in $\sigma_2(\omega)$ (panels c.1, c.3) and a gapped $\sigma_1(\omega)$ (panels c.2, c.4), both below and above $T_C$. At even shorter pump wavelengths (400 nm) both the imaginary part of the conductivity $\sigma_2(\omega)$ (panels d.1, d.3) and the real part $\sigma_1(\omega)$ (panels d.2, d.4) increase considerably at all time delays, indicating a simultaneous enhancement in interlayer tunneling *and* an increase in the incoherent transport.

The observations reported above are better understood by time-delay ($\tau$) dependent measurements. The combined coherent and incoherent response is well captured in the energy loss function $-\text{Im}[1/\tilde{\varepsilon}(\omega,\tau)]$, which is displayed for both temperatures above and below $T_C$ in figure 5. The loss function exhibits a peak where $\tilde{\varepsilon}$ crosses zero, that is, at the frequency of the plasma edge (see figure 2). The width of the loss function reflects instead the scattering rate or, equivalently, the inverse coherence length for superconducting tunneling. The same features reported above are identified in these plots, that is, the progressive increase of the loss-function peak frequency and incoherent broadening which sets in at earlier delays for shorter wavelength excitation.

By fitting the loss function and the other transient optical properties with a Drude model, both the screened plasma frequency $\widetilde{\omega}_p$ (loss function peak frequency) and the scattering time $\tau_s$ (inverse width of the loss function peak) are extracted for all measured temperatures, pump wavelengths and time delays. These two quantities are plotted in figure 6a and 6b at $\tau$ = 1.5 ps pump-probe delay as a function of excitation wavelength. The following picture emerges: interlayer tunneling is enhanced most effectively at higher pump photon energies, but the transient phase also relaxes most rapidly and drives incoherent processes after shorter wavelength photo-excitation.

The observation that transient interlayer coupling is strengthened most effectively when striped cuprates are excited at short wavelengths is highly surprising. The most likely explanation is that strong electronic correlations are dominant in stabilizing charge order in the first place[31]. Note that a conventional interpretation of competing charge-density-wave and superconducting order posits that the two orders interact on energy scales commensurate with the ordering temperatures (~10 meV). This is not the case here, where interactions on high energy scale compete and cooperate to provide order at far lower energies. More generally, our work shows how the use of light to switch between different symmetries in complex materials can provide microscopic information on the stability of individual orders, and complement linear spectroscopies in important ways.

**Acknowledgments**

The research leading to these results has received funding from the European Research Council under the European Union's Seventh Framework Programme (FP7/2007-2013)/ERC Grant Agreement No. 319286 (Q-MAC), and from the German Research Foundation (DFG-SFB 925). Work at Brookhaven was supported by the Office of Basic Energy Sciences, Division of Materials Sciences and Engineering, U.S. Department of Energy under Contract No. DE-SC00112704.

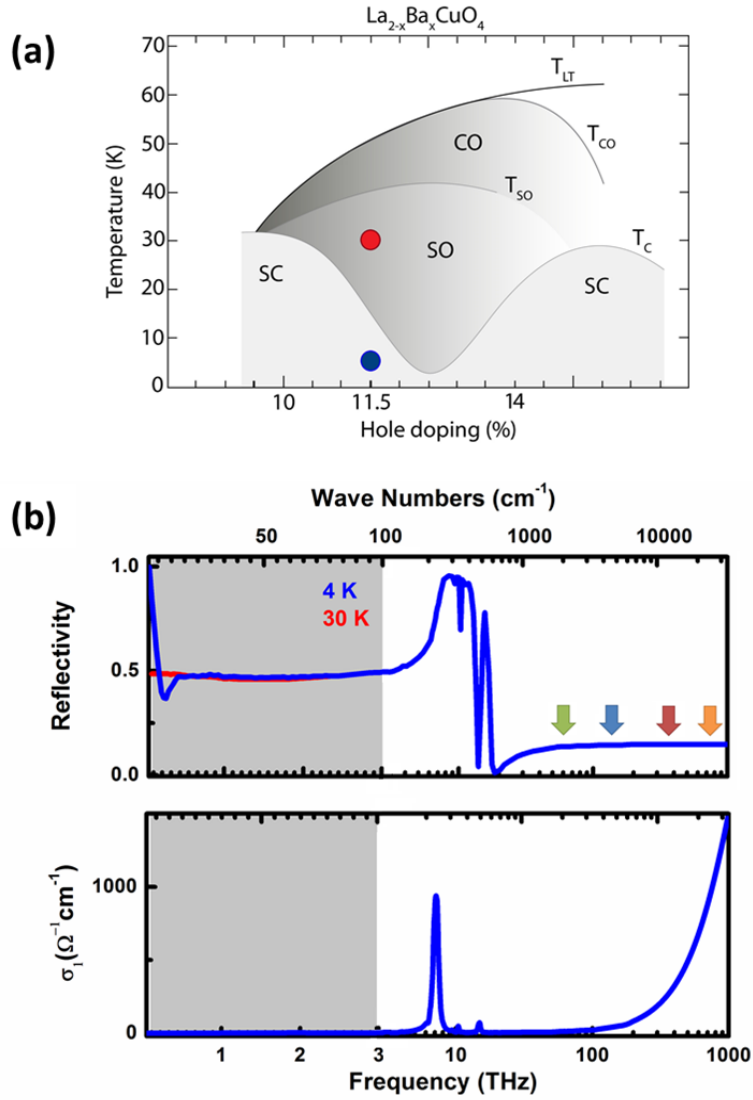

**Figure 1.** (a) Phase diagram of $La_{2-x}Ba_xCuO_4$ as a function of temperature and doping, as determined in Ref. 25. SC, SO, and CO indicate the superconducting, spin-order, and charge-order states, respectively, with $T_C$, $T_{SO}$, and $T_{CO}$ being the corresponding transition temperatures. $T_{LT}$ represents the structural transition temperature. The circles indicate the different temperatures for which the current experiment has been carried out: T = 4 K (blue) and T = 30 K (red). (b) *c*-axis optical properties of $La_{1.885}Ba_{0.115}CuO_4$ at equilibrium: Frequency-dependent reflectivity (upper panel) and real part of the optical conductivity, $\sigma_1(\omega)$, (lower panel). The spectral region probed in the current experiment is highlighted in grey. The arrows indicate the pump wavelengths used for excitation: 5 µm (green), 2 µm (blue), 800 nm (red), and 400 nm (orange).

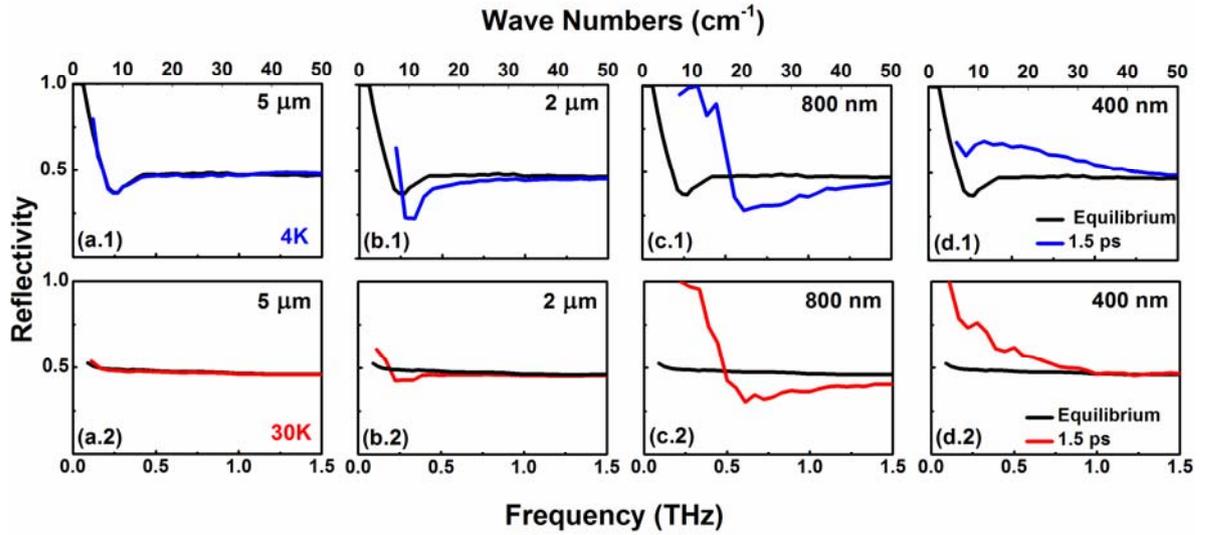

**Figure 2.** Frequency-dependent reflectivity of the photo-stimulated $La_{1.885}Ba_{0.115}CuO_4$ layer, measured at equilibrium (black lines) and 1.5 ps after excitation with different pump wavelengths. Spectra are shown in a limited spectral region (to highlight pump-induced changes), at two different temperatures: $T$ = 4 K (blue) and $T$ = 30 K (red). All data were taken at a pump fluence corresponding to ~$2 \cdot 10^{20}$ photons/cm$^3$.

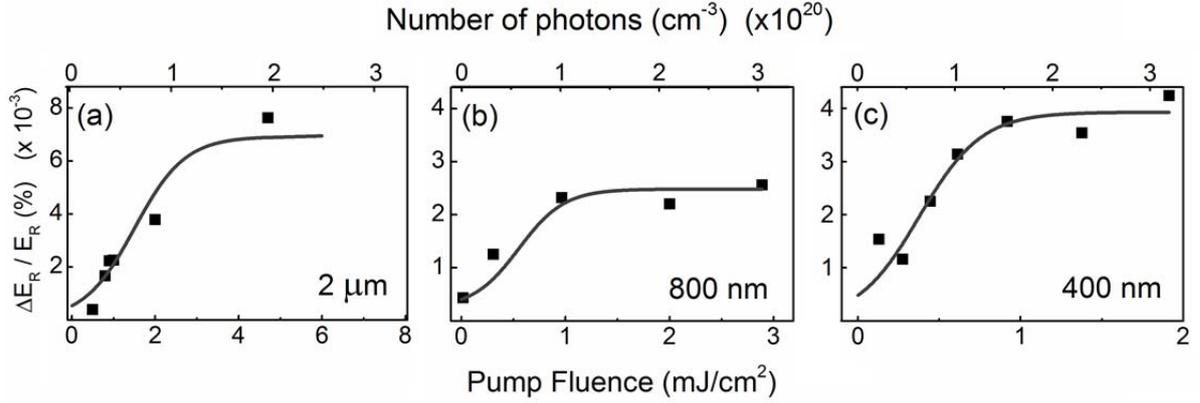

**Figure 3.** Differential time-domain transient ΔE$_R$/E$_R$ measured at the THz electric field peak, 1.5 ps after photo-excitation, plotted as a function of pump fluence for different excitation wavelengths. The black lines are sigmoid function fits, returning threshold fluences of 3.0 mJ/cm², 1.1 mJ/cm² and 0.75 mJ/cm² for 2-μm, 800-nm and 400-nm excitation wavelength, respectively. On the top horizontal scale the fluence is expressed in terms of total number of absorbed photons per unit volume, returning a saturation value of ~10$^{20}$ photons/cm³, independent of pump wavelength.

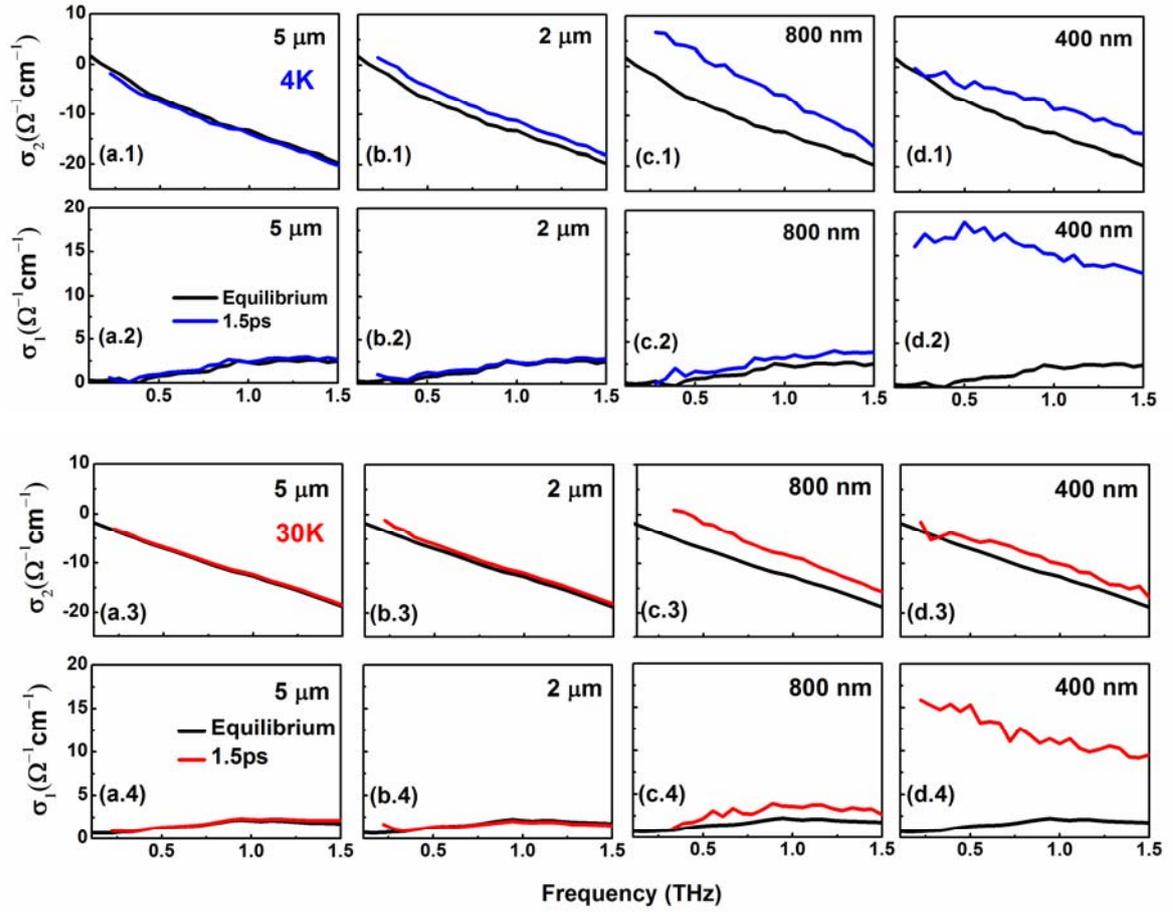

**Figure 4.** Complex optical conductivity of $La_{1.885}Ba_{0.115}CuO_4$ at equilibrium (black) and 1.5 ps after excitation with different pump wavelengths (colored). All data were taken at a pump fluence corresponding to ~2·10$^{20}$ photons/cm$^3$, both below (blue) and above (red) $T_C$.

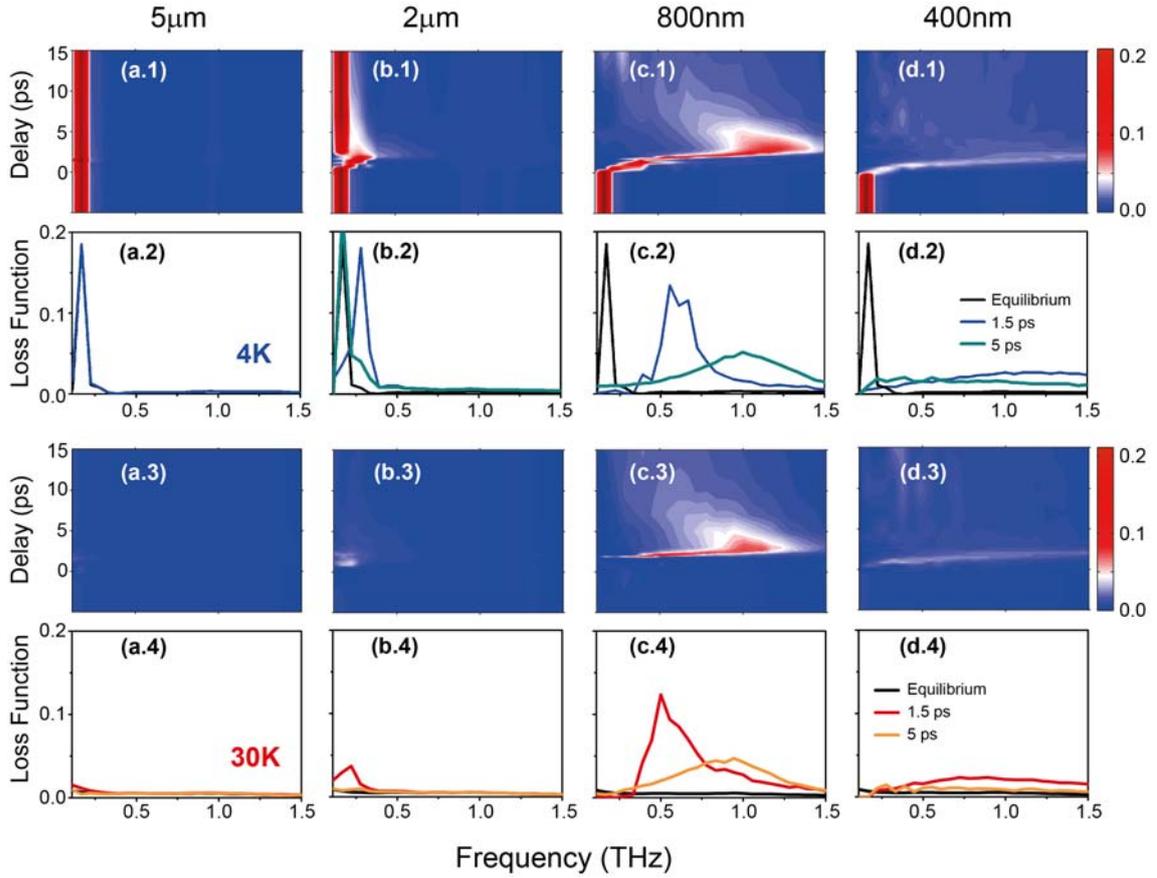

**Figure 5.** Frequency-dependent Energy Loss Function $[-\text{Im}(1/\tilde{\varepsilon})]$ of $La_{1.885}Ba_{0.115}CuO_4$ for different excitation wavelengths, as a function of pump-probe delay, measured below (panels 1-2) and above (panels 3-4) $T_C$. Color plots show the light-induced dynamical evolution of the loss function, while selected line cuts are reported at negative (black), + 1.5 ps (blue and red) and + 5 ps (cyan and orange) time delay. All data were taken with a pump fluence corresponding to ~$2\cdot10^{20}$ photons/cm$^3$.

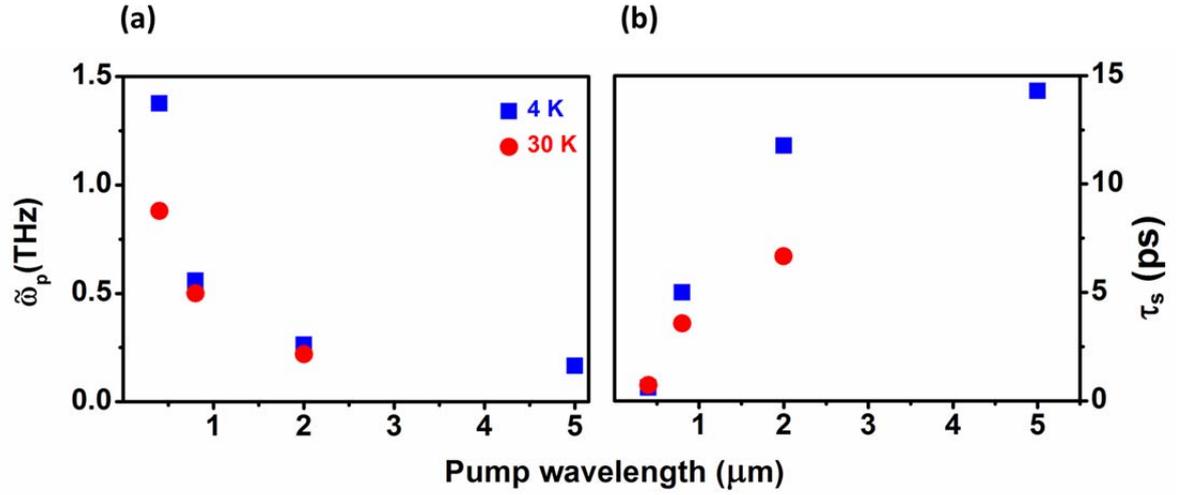

**Figure 6.** (a) Screened plasma frequency and (b) scattering time of the transient state, displayed as a function of pump wavelength, for temperatures below (blue) and above (red) $T_C$. All values have been extracted via Drude fits to the spectra measured at τ = 1.5 ps with a pump fluence corresponding to ~2·10$^{20}$ photons/cm$^3$. Direct estimates of $\widetilde{\omega}_p$ and $\tau_s$ were also obtained from the loss function peak frequency and inverse width, in perfect agreement with those extracted from the fits.

[24] Light polarized out-of-plane does not couple to quasiparticle excitations in two-dimensional superconductors, and is expected to affect the superconducting condensate only weakly in the quasi-2D case of cuprates. On the other hand, because of the peculiar arrangement of charge order in LBCO, in which parallel stripes in next-to-neighbor planes are shifted by π, coupling to the charge stripes and dipole activity is expected for this polarization.

[25] M. Hücker, M. v. Zimmermann, G. D. Gu, Z. J. Xu, J. S. Wen, Guangyong Xu, H. J. Kang, A. Zheludev, and J. M. Tranquada, "Stripe order in superconducting $La_{2-x}Ba_xCuO_4$ (0.095⩽$x$⩽0.155)", Phys. Rev. B **83**, 104506 (2011).

[26] The THz-frequency (0.15 - 3 THz) reflected fields measured here below and above $T_C$ were combined with the broadband (up to ~1,000 THz) reflectivities reported in Ref. 27. By applying Kramers-Kronig transformations, a full set of equilibrium optical properties (*i.e.* complex dielectric function, complex optical conductivity, complex refractive index) could be determined at both temperatures investigated in the present experiment.

[27] C. C. Homes, M. Hücker, Q. Li, Z. J. Xu, J. S. Wen, G. D. Gu, and J. M. Tranquada, "Determination of the optical properties of $La_{2-x}Ba_xCuO_4$ for several dopings, including the anomalous *x* = 1/8 phase", Phys. Rev. B **85**, 134510 (2012).

[28] The pump-induced changes in amplitude and phase of the reflected THz electric field were measured at different pump-probe delays. These "raw" reflectivity changes were only ≲1% due to the frequency-dependent mismatch between the penetration depth of the THz probe (~ 50 – 500 μm) and that of the pump field (0.1, 0.4, 2.5, and 10 μm for $\lambda_{pump}$ = 400 nm, 800 nm, 2 μm, and 5 μm, respectively). Such mismatch was taken into account by modeling the response of the system as that of a homogeneously photo-excited thin layer on top of an unperturbed bulk (which retains the optical properties of the sample at equilibrium). By calculating the coupled Fresnel equations of such multi-layer system [M. Dressel and G. Grüner, Electrodynamics of Solids, Cambridge University Press, Cambridge (2002)], the transient optical response (reflectivity, energy loss function, complex optical conductivity) of the photo-excited layer could be derived. A detailed explanation of this procedure can also be found in Ref. 29. The calculated optical properties were then compared with those obtained by treating the excited surface as a stack of thinner layers with a homogeneous refractive index and describing the excitation profile by an exponential decay (see Ref. 30). The results of the two models were found to be in agreement within a few percent.

[29] W. Hu, S. Kaiser, D. Nicoletti, C.R. Hunt, I. Gierz, M. C. Hoffmann, M. Le Tacon, T. Loew, B. Keimer, and A. Cavalleri, "Optically enhanced coherent transport in $YBa_2Cu_3O_{6.5}$ by ultrafast redistribution of interlayer coupling", Nat. Mater. **13**, 705 (2014).

[30] S. Kaiser, C. R. Hunt, D. Nicoletti, W. Hu, I. Gierz, H. Y. Liu, M. Le Tacon, T. Loew, D. Haug, B. Keimer, and A. Cavalleri, "Optically induced coherent transport far above $T_C$ in underdoped $YBa_2Cu_3O_{6+\delta}$", Phys. Rev. B **89**, 184516 (2014).